\documentclass[aps,prd,10pt,twocolumn,showpacs,preprintnumbers,amsmath,amssymb,nofootinbib]{revtex4-1}

\usepackage{natbib} 
\usepackage{color}
\usepackage{amsmath}
\usepackage{epsfig}
\usepackage{array}
\usepackage{hyperref} 

\usepackage{natbib}
\usepackage{graphicx}

\begin{document}

\title{A Hybrid Approach for Using Programming Exercises in Introductory Physics}

\author{Orban, C.$^*$}
\affiliation{Department of Physics, The Ohio State University, Columbus, OH, 43210}
\author{Teeling-Smith, R. M.}
\affiliation{University of Mt. Union, Alliance, OH, 44601}
\author{Smith, J. R. H.}
\affiliation{Department of Physics, The Ohio State University,Columbus, OH, 43210}
\author{Porter, C. D}
\affiliation{Department of Physics, The Ohio State University,Columbus, OH, 43210}

\email{orban@physics.osu.edu}

\date{\today}

\begin{abstract}

Incorporating computer programming exercises into introductory physics is a delicate task that involves a number of choices that may have an effect on student learning. We present a ``hybrid" approach that speaks to a number of common concerns regarding cognitive load which arise when using programming exercises in introductory physics classes where many students are absolute beginner programmers. This ``hybrid" approach provides the student with a highly interactive web-based visualization, not unlike a PhET or Physlet interactive, but importantly the student is shown only the subset of the code where the initial conditions are set and the system variables are evolved. We highlight results from a coding activity that resembles the classic game Asteroids. The goals of this activity are to show how a simple 1D code can be modified into a 2D code, and to reinforce ideas about the relationship between force, velocity, and acceleration vectors. Survey results from four semesters of introductory physics classes at the Ohio State University's Marion campus, in which a high percentage of the students are weak or absolute beginner programmers, provide evidence that most students can complete coding tasks without severe difficulty. Other survey results are promising for future work where conceptual learning will be assessed in a direct way using metrics like the Animated Force Concept Inventory \cite{Dancy2006}.  The exercise highlighted here and others from our group are available for general use at \url{http://compadre.org/PICUP}. 
\end{abstract}

\maketitle

\section{Introduction}
\label{sec:intro}

The need to incorporate programming content into introductory physics is widely appreciated by the academic community \cite{Fuller2006}. By some estimates, \emph{at least 70\%} of new STEM jobs in the US will require computer programming skills \cite{labor2014} and in the sciences computer programming skills have become an essential part of many disciplines. In response to these shifts, groups like code.org and the ``hour of code" have brought coding tutorials to wider and younger audiences \cite{code}. These groups also influenced federal education legislation in the US. In particular, the Every Student Succeeds Act (ESSA), which was signed into law in December 2015, designates computer science as a ``core subject". This is a significant change that places computer science on the same level as english and mathematics \cite{coresubject}. The 2017-2018 school year was the first school year that this legislation was fully implemented.  Yet, for physics instruction, and perhaps even more generally, the task of re-imagining STEM courses with computer science as a crucial element is still far from complete. Although there are a number of universities that use coding activities in physics with vpython \cite{Chabay_Sherwood2008}, and there exists significant research into using these activities in calculus-based introductory physics \cite{Caballero_etal2012}, vpython exercises (or coding using some other software framework) are much less often used in algebra-based physics and at the high school level. 

We were able to find two studies that reflect the difficulty of using coding activities in algebra-based physics at the high school level\footnote{The open source ebook by \citet{Titus_Esquembre2016} is also notable but there seems to be no published studies examining its appropriateness for various grade levels.}.  \citet{Aiken_etal2013} describes a masters degree project by a high-school physics teacher who worked for two years to develop a vpython curriculum for a 9th grade high school physics class and found that only one third of the class successfully completed the exercises. \citet{Aho2014} describes coding activities developed for a high school classroom that use the R programming language. Although they are not very specific in stating precisely what fraction of the students struggled with the exercises, they do indicate that a significant number of students needed extra time outside of class to complete the activities, and these students frequently needed extra practice to learn the R syntax. To mitigate this in future work, \cite{Aho2014} proposes to set aside a week-long R programming tutorial for the students at the beginning of the year, which is a luxury most high school teachers do not have. The indications from \citet{Aiken_etal2013} and \citet{Aho2014} underscore the need to develop a curriculum that adds programming into \emph{algebra-based} physics with a higher success rate.

\begin{figure*}
\includegraphics[width=2.4in]{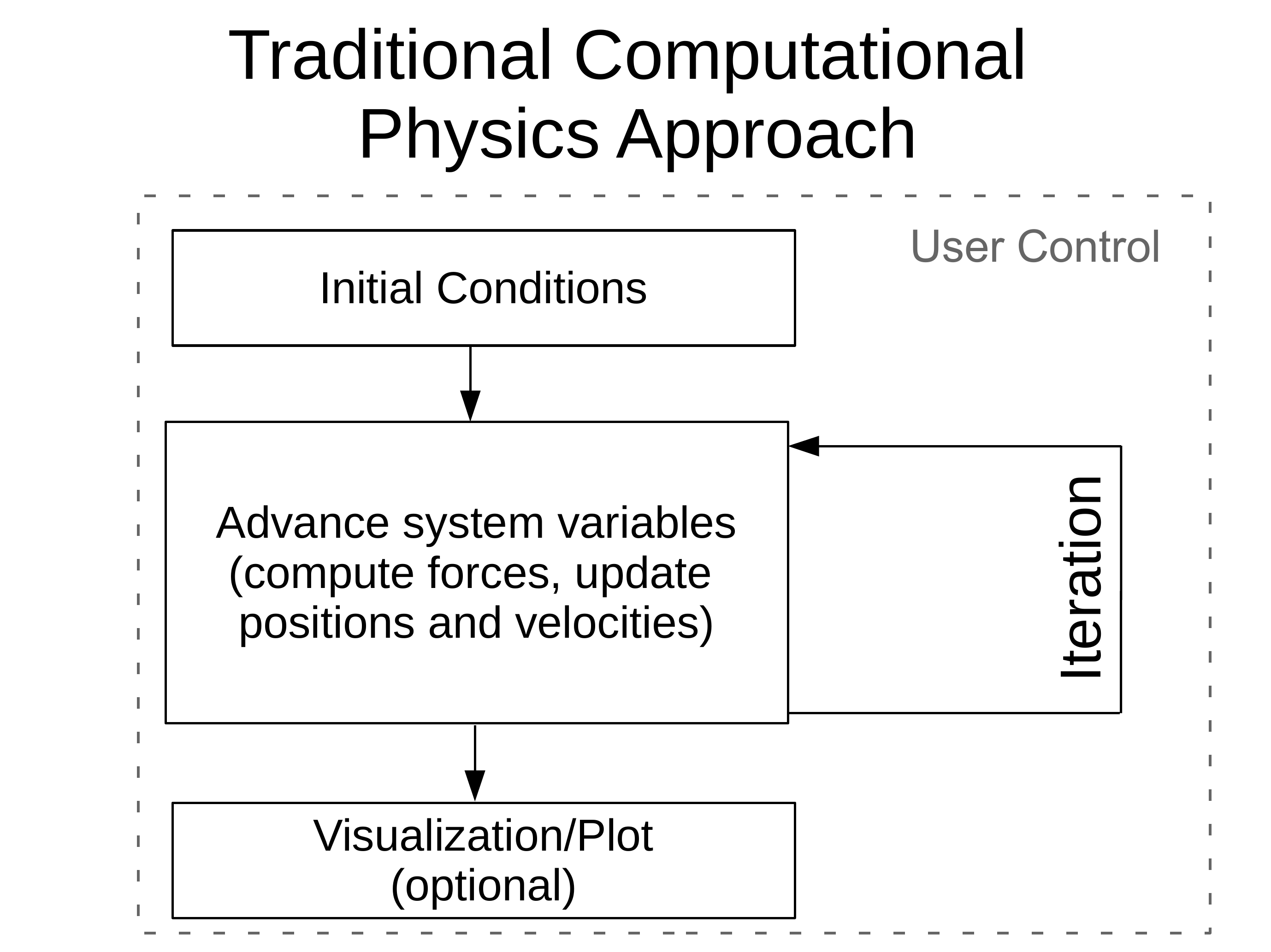}\includegraphics[width=2.4in]{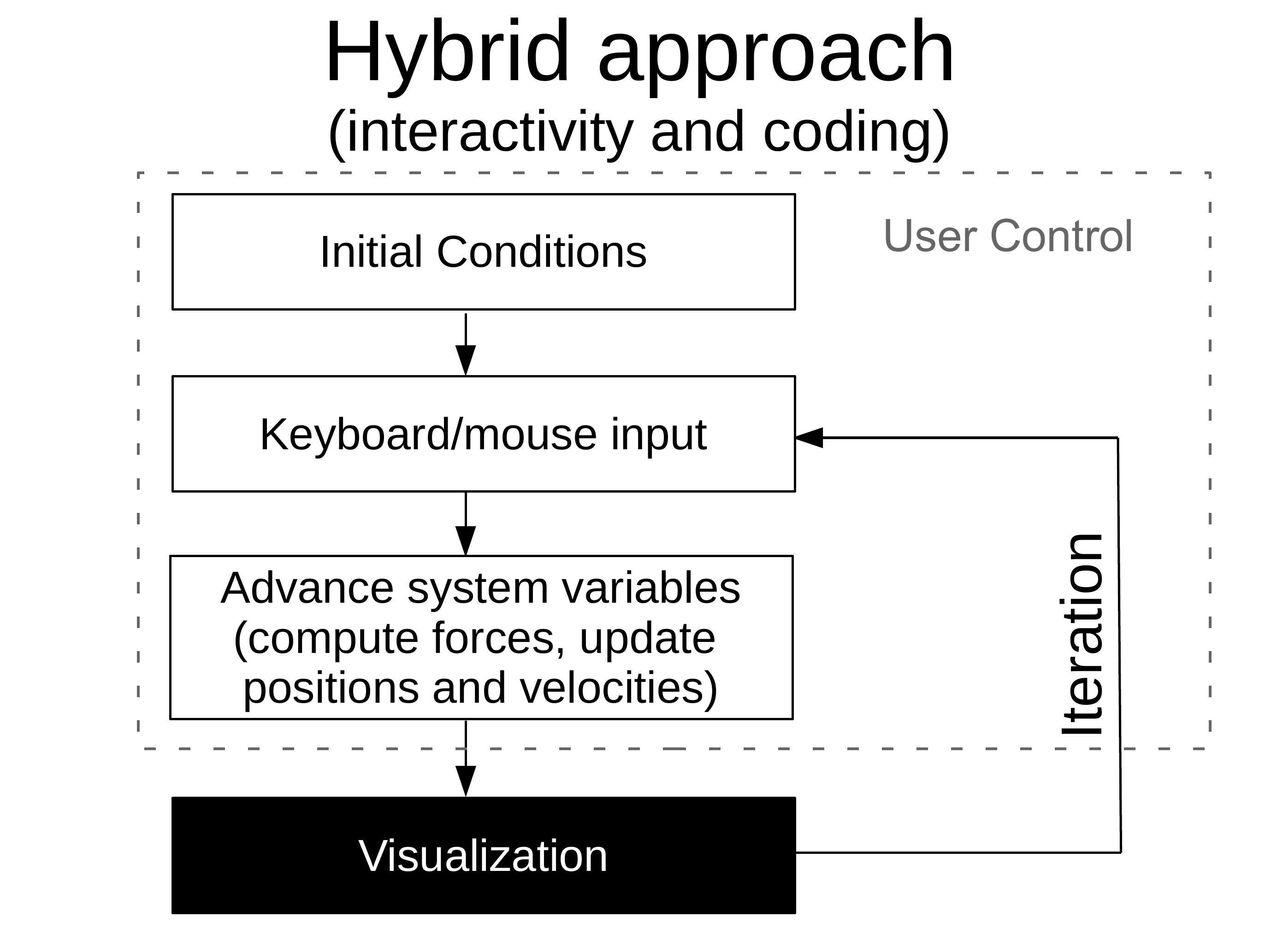}\includegraphics[width=2.4in]{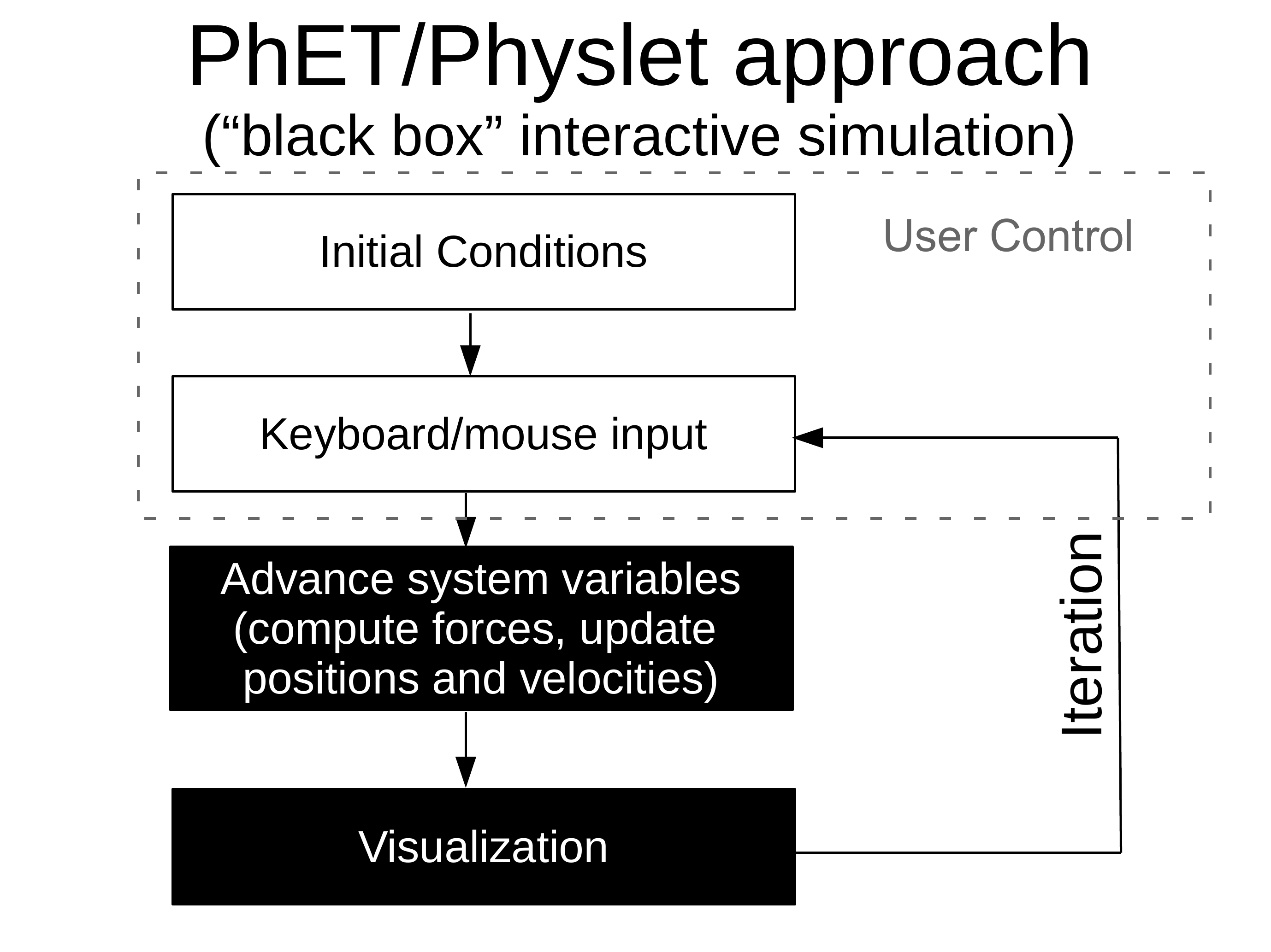}
    \caption{Illustrations of different approaches to computationally-enriched physics content. The left panel illustrates the typical structure of a code in a traditional intermediate-to-advanced level physics-major computational physics course, emphasizing that the student has control over essentially the entire code. The right panel shows the typical structure of a web interactive in which students interact with a visualization but do not see or have any control of the underlying code. The center panel shows a hybrid approach with a high degree of interactivity but the students do see and potentially modify the parts of the code that advance system variables, even though code related to visualization remains fixed and invisible to the student.}
    \label{fig:approach}
\end{figure*}

As will be discussed, we use a javascript-based language called p5.js, which was designed to be a text-based (as opposed to block-based) language with a gentle learning curve for absolute beginner programmers. In principle, the exercises we describe here could be reproduced in vpython (or some other language) and used in a similar way. While there are clearly advantages and disadvantages to both vpython and javascript, the comparison of the two is not the subject of this paper. Instead we wish to emphasize the need for coding activities that would be appropriate for an algebra-based physics classroom. As discussed later, an important way to judge the appropriateness of these activities is the perceived difficulty of students who complete coding activities.

The left hand panel in Figure~\ref{fig:approach} illustrates what we describe as the ``traditional" computational physics approach that appears in an intermediate-to-advanced physics-major computational physics course, or in a physics-major course that has been re-tooled to include significant computational content. In this setting, the student is given complete control of the computer program, including the advance of variables (which may involve specifying forces and advancing positions and velocities, or it may involve the evolution of abstract quantities like wave functions). If visualization is needed, the student is typically given full control of a plotting program. Although there may be some template that the student is given and other advice may be provided, overall, the student has a high level of control. The drawback for this approach is that significant class time is often required for students to familiarize themselves with coding practices. Given the time constraints of a typical algebra-based college physics course, or high school physics course, this approach is in-feasible for most instructors.

The right hand panel in Figure~\ref{fig:approach} describes interactive physics simulations in which the students do not see the code. This approach is extensively used by the PhET collaboration \cite{PhET} and by the ``Physlet" physics community \cite{physlet}, and many studies have shown its utility for teaching scientific concepts \cite[e.g.][]{Perkins_etal2006,Podolefsky_etal2010}. Largely for this reason, PhET and similar activities have been put into widespread use.

The central panel in Figure~\ref{fig:approach} outlines the ``hybrid" approach that we adopt in this paper in which the student does see and potentially modifies the code that evolves system variables (which is similar to the traditional approach), and there is some kind of interactive visualization that is produced in which the simulation responds to user input (which is similar to the PhET/Physlet approach). However there are still aspects of the code, particularly related to visualization, that the student does not see in order to substantially reduce the cognitive load \cite{Jong_2010,cogload} by shortening the length and minimizing the complexity of the program. The intention is to remove ``extraneous cognitive load" associated with the graphical user interface among other things, and focus on the aspects of the code that directly determine or update physical quantities. Our assumption is that the ``intrinsic cognitive load" of setting and updating the physical quantities using the target concepts and relationships is within students' abilities. As will be illustrated in this paper, the portion of the code with which the students interact can be concise both textually and conceptually, and still produce interesting game-like interactives that emphasize kinematic and diagrammatic concepts like force, acceleration, velocity, and their vector representations.

To provide some comparison to other works in the literature, there may also be some overlap with our approach and that of \cite{kordakai2010}, who describes a graphically enriched coding environment for teaching computer science and outlines how their activities align with the educational theories of various authors. The Netlogo project \cite{netlogo} is another comparable effort which borrows from earlier efforts to incorporate programming into schools, but we are not familiar enough with how Netlogo activities are used in introductory physics to say more than this.

Although there are exceptions \cite{Taub_etal2015,Weintrop_etal2016,Titus_Esquembre2016,netlogo}, interactive activities where students key-in commands and ``play" their code like a video game, are typically not a part of programming exercises at the introductory level. In the Matter \& Interactions curriculum that integrates vpython into calculus-based physics \cite{Chabay_Sherwood2015}, many of these programs, such as the three-body gravitational simulation or the 3D pendulum \cite{Chabay_Sherwood2008}, are designed for the student to perform coding tasks and then passively watch the execution of the program (except perhaps for changing the perspective). And while there are a large number of exercises currently available on the AAPT's Partnership for the Integration of Computation into Undergraduate Physics (\href{http://compadre.org/PICUP}{compadre.org/PICUP}), only a few of them involve a high level of interactivity as the program is running. 
Our hypothesis is that this interactive, game-like approach with a concisely-written code will create a fun and approachable experience for students who might otherwise find a programming task to be intimidating, making it an ideal choice for engaging students in introductory courses.

This paper is only the beginning of a research effort to validate this hypothesis. We will describe a set of computer programming activities designed for absolute beginner programmers in first-semester introductory physics (mechanics) classes, that were used during four semesters at Ohio State's Marion campus. Survey results will be presented that examine student perceptions from completing the first exercise, and probe the percentage of weak or absolute beginner programmers in the classroom. 

Although there is good work in the literature describing how numerical exercises can be connected with laboratory activities \cite[e.g.][]{Serbanescu_etal2011}, we consider this out of scope for the present work. The javascript-based language p5.js does have capabilities to interact with Arduino circuit boards, making this an interesting possibility for future work.

\section{Overview of Programming Activities for Mechanics}

In a semester course of introductory physics at Ohio State University (OSU) at the regional campus in Marion, we include six required programming activities and a seventh activity that is optional or extra credit. In most other ways, the course is identical to the same course on the Columbus campus. The official description of this course is calculus-based physics I, but on all OSU campuses students only need to be concurrently enrolled in calculus in order to take the course, and as a result the calculus content in the course is rather limited. Moreover, the students at OSU's regional campuses are less prepared than their peers on OSU's Columbus campus. During the data gathering, incoming OSU Marion students had an average ACT score near 22 (in 2014 \cite{osumarion2014} and 2015 \cite{osumarion2015}) or 22.5 (in 2016 \cite{osumarion2016}) whereas students admitted directly to the Columbus campus over the same time span had an average ACT score close to 29 \cite{osucolumbus}. The limited calculus in the course and the comparably poor ACT performance of the students make interesting venue for integrating programming exercises into introductory physics with the end goal of creating a curriculum that might succeed in the high school physics classroom.

Each activity is designed to take about an hour to complete. Students are not explicitly assigned to groups or pairs, but the classroom setup involves six tables of four, so students will tend to collaborate on the activities and this is not discouraged. To date, about 125 students from OSU's regional campus in Marion have completed the exercises mentioned below. The activities are graded on the completion of required steps.

All of the exercises illustrate the velocity, acceleration and force vectors. The first exercise gives the student much of the code that they will need, only asking them to make small, guided modifications, which we will describe in the next section. All of the exercises build off of each other in a way that would make it hard for a student to start in the middle of the sequence.  Additionally, all of the exercises contain ``challenges" that encourage the student to develop some functionality that often adds an interesting element to the game. The list of exercises is as follows:
\begin{enumerate}
    \item Planetoids (similar to the classic game ``Asteroids")
    \item Lunar descent (similar to the classic game ``Lunar lander")
    \item Bellicose birds (similar to the popular game ``angry birds")
    \item Planetoids with momentum
    \item Planetoids with torque
    \item Planetoids with a spring (harmonic motion)
    \item Extra credit: Bellicose birds with energy
\end{enumerate}

This sequence is designed to accompany a typical physics course on classical mechanics where momentum is not introduced until mid-way through the course, followed by concepts of torque and, later, harmonic motion. The ``Bellicose birds with energy" exercise is made available to students in the middle of the course when energy is introduced, but this exercise is more difficult than the others because it is the only exercise that deals explicitly with the integration scheme. For simplicity, all the exercises adopt Euler-Cromer integration \cite{Cromer1981} except for ``Bellicose birds with energy", which describes the trapezoidal method in terms that an algebra-based physics student should be able to understand.

In this paper we provide a rather extensive description of the first exercise (``Planetoids") including student survey responses. This section provides a context to this exercise since essentially all the exercises listed above are derived from this ``Planetoids" exercise.  These activities will be described in detail in later work. We will only add here that some of these additional activities use a graphing system to plot various relevant physical quantities (such as velocity) over time in the bottom right corner of the screen.



\begin{figure*}
    \centering
    \includegraphics[width=5in]{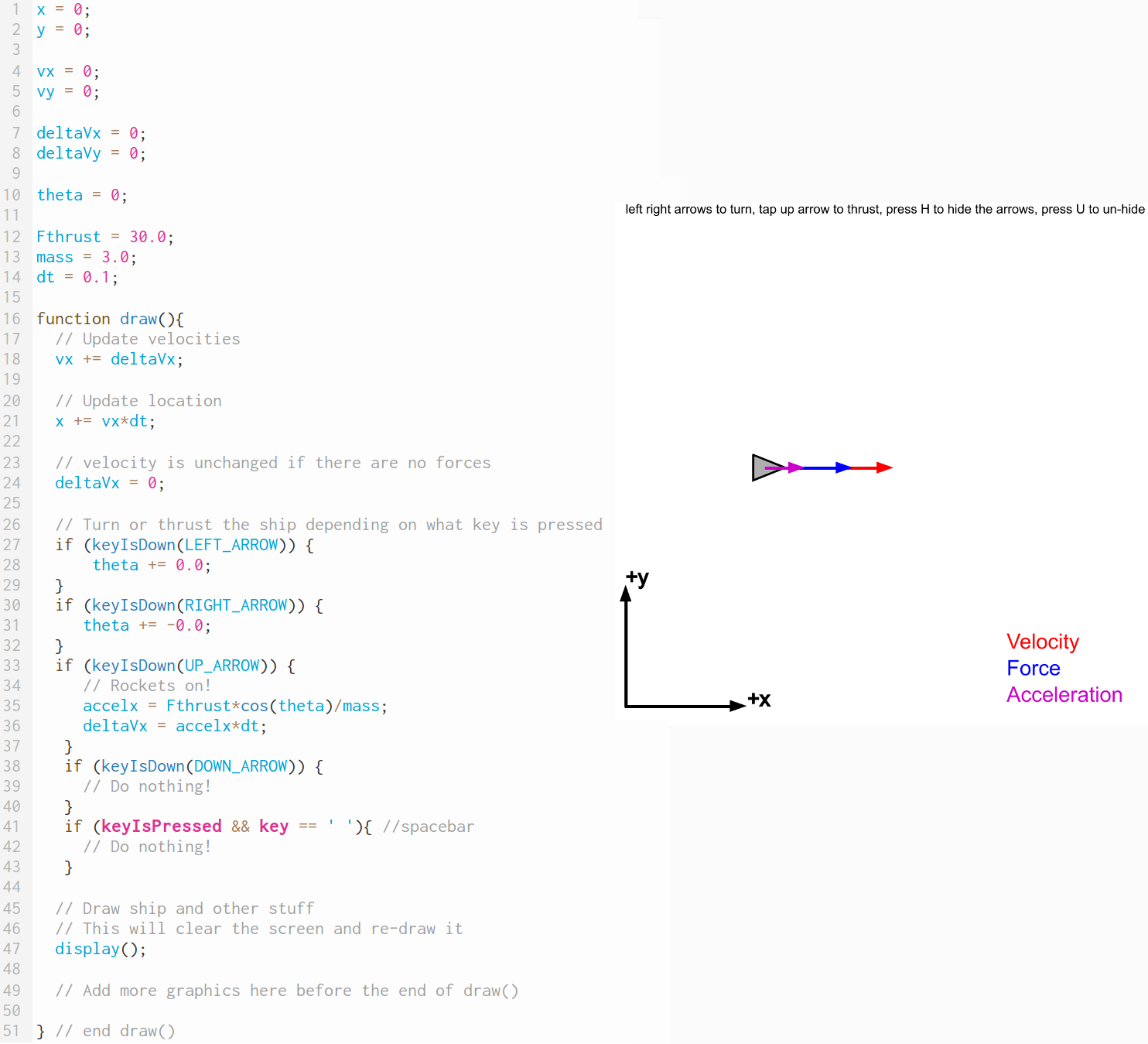}
    \caption{The code (left) and corresponding interactive (right) that the student sees at the beginning of the first exercise. This code is written with the processing javascript library p5.js. As a result, the code has a C/C++ like syntax except that draw() replaces main() and draw() is run 60 times per second until the user stops the program. The interactive (right) shows a ship traveling towards the right with constant velocity indicated by a red velocity vector. On the left panel the student sees about 50 lines of code, but about half of these lines are spaces or comments.}
    \label{fig:planetoids}
\end{figure*}

\section{The Planetoids Game}

The choice of an ``Asteroids"-like game for the first activity is intentional. A natural environment for illustrating Newton's laws is free space, away from any sources of gravity. We are not unique in using this situation as a starting point. \cite{white1984} found learning gains from students interacting with a video game with a similar premise, and no doubt other authors adopt a similar approach. The advantage of this environment is that objects in motion will continue with the same velocity, moving in a straight line, unless a force is acts upon them. The classic game ``Asteroids" illustrates this well with a ship that drifts through free space, except when its rockets fire to avoid asteroids that are also drifting through free space. The net force is either zero, or constant in the direction the ship is pointing.

\subsection{Learning Goals}
\label{sec:learning}

The learning goals for this exercise are as follows:

\begin{enumerate}
    \item Understand how to convert a simple 1D code into a 2D code
    \item Understand how force, velocity and acceleration vectors relate to the motion of a ship traveling in free space
\end{enumerate}

The list above is intentionally short because we do not expect that during this 1-2 hour activity that the student will be able to absorb the subtleties of computational thinking \cite{Weintrop_etal2016}
or become proficient with the javascript-based coding framework to the point where they can comfortably make numerous modifications to the code. In the following subsections we discuss how the activity is structured to reinforce the two learning goals mentioned above, and we point out various difficulties that students often have.  We discuss additional learning goals and extensions to the activity in later sections.

\subsection{Structure of the Program and Design Choices}

Figure ~\ref{fig:planetoids} shows what the student sees at the beginning of this exercise. Initially, the ship can only move in the $x$ direction and the first task is to allow the ship to rotate when the user presses the left and right arrow buttons by changing the value of $\theta$. It is worth commenting on Fig.~\ref{fig:planetoids} in detail because even at this stage there are a number of choices that have been made that could affect student learning. One important choice made to simplify the cognitive load for the student is to ``hide" a significant amount of code in the \texttt{display()} function. In this example, there are about three times more lines of code defining the \texttt{display()} function than the $\approx 50$ lines of code that the student sees and modifies\footnote{We attempted to re-create this exercise in vpython using glowscript.org but found that (at least currently) there is no way of setting up a second page of code where subroutines can be defined without being in plain view by the student, which is a barrier to implementing this ``hybrid" approach. This may or may not be a limitation with other browser-based python development environments like trinket.io or jupyter.org}.

Another important choice is to hide the variable types. There are no \texttt{float}, \texttt{int}, \texttt{double} or \texttt{var} declarations used to initialize the variables. Instead, variables are implicitly declared to be floating point decimals and the number of characters that the student sees is minimized. This syntax is essentially the same as used in Matlab, which is a popular language for absolute beginner programmers. We use the processing javascript library p5.js for these exercises and as a result the code shown in Fig.~\ref{fig:planetoids} is javascript which does not produce an error for missing variable types. A possible drawback of postponing the discussion of variable types is that the difference between global and local variables is not explained at this stage. Students may not realize that \texttt{accelx}, which is only used and defined inside of an \texttt{if} statement, is a local variable while \texttt{deltaVx} is a global variable, but this is unlikely to cause a problem at this stage. Our philosophy is to explain subtleties like these in the step-by-step tutorial only if absolutely necessary for completing a particular exercise.

The structure of the program in Fig.~\ref{fig:planetoids} is an important choice that may affect student learning. The sections of the code are as follows:
\begin{enumerate}
\item Variable initializations 
\item the \texttt{draw()} function -- velocity and position advance
\item the \texttt{draw()} function -- keyboard inputs
\item the \texttt{draw()} function -- \texttt{display()} function followed by other user-defined graphics
\end{enumerate}
It is understood that the \texttt{draw()} function is run many times per second so that after the \texttt{display()} function is executed the program will go back to the beginning of \texttt{draw()} and advance the velocity and position again and go through the whole sequence again until the user presses stop\footnote{An optional ``Hello world" activity demonstrates that adding code to write a simple message to the browser console while inside of the \texttt{draw()} function will result in that message being written many times over because the \texttt{draw()} function is being run many times per second.}. Because \texttt{draw()} is being run again and again, one could easily change the sequence so that, for example, the \texttt{display()} function would be first and the velocity and position advance would be last. The drawback of this approach is that when the student parses the code for the first time they would see the physics content of the code \emph{last}. In a physics course, our primary interest lies in directing the students' attention to how the physics content, such as $d=vt$ for example, is implemented in the code, with discussions of the programming concepts such as syntax, variable types, and the structure of the algorithm being supplementary to that.

Following the physics section there is a line of code \texttt{deltaVx = 0} ($\Delta v_x = 0$) which is accompanied by a comment ``velocity is unchanged if there are no forces". This is just a restatement of Newton's first law in a form that a computer can understand. Following this, the program checks if the user is pressing certain buttons on the keyboard.

The drawback to this physics-first, keyboard commands later approach is that the student may not fully appreciate that the program holds on to the global variable \texttt{deltaVx}, which is determined from the keyboard command section, only using it again at the beginning of the \emph{next} iteration of \texttt{draw()}.

In the written step-by-step directions, the user is asked to put non-zero values in the section of the keyboard input section that changes the angle of the ship. Then the student is asked to enable motion in the $y$ direction by imitating the code for advancing the velocity and position in the $x$ direction. Finally, the student is asked to determine the correct change in velocity due to a constant force (thrust) in the $y$ direction. This involves realizing that while $\cos \theta$ gives the component of the force oriented in the $x$ direction, one must use $\sin \theta$ to obtain the component of the force in the $y$ direction.  Students are given a hint that it is either a cosine, sine, or tangent function that gives the correct behavior.

At each step in the tutorial, the student can click links to see and interact with how the program should work at a particular stage, but without seeing the source code for the completed step. This is an important capability that gives the student instant guidance on whether they have completed a particular programming task correctly, leaving the instructor more time to spend on subtle issues.

Common mistakes that students make include forgetting to set \texttt{deltaVy = 0}, in which case the ship accelerates uncontrollably in the $y$ direction. Students rarely self-diagnose this issue because the ship appears to behave correctly if the thrusters are repeatedly fired and it is only when the student stops firing the thrusters that the uncontrollable acceleration becomes obvious. When students interact with the correct version of the program (as described in the previous paragraph) they should notice this difference in behavior but the problem is subtle enough that this problem is easy to miss.

Another frequent mistake is that students tend to do a quick copy paste of the acceleration code for the $x$ direction to the $y$ direction without changing the trigonometric function from cosine to sine. This causes $\Delta v_y = \Delta v_x$ and as a result the ship only travels on a diagonal line regardless of the angle $\theta$. Students have an easier time self-diagnosing this issue because the problem is easy to see and they are told that the trigonometric function in the line of code that determines $\Delta v_y$ should be either a cosine, sine, or tangent.

\subsection{Challenges}

Students must also implement 1-2 ``challenges". The challenges in this exercise include creating ``planetoids" (a word play on the astronomical term planetesimals) that drift across the screen using the \texttt{drawEllipse()} function and adding reverse thrusters when the down arrow is pressed (which can be done by copying the code from the up arrow and adding minus signs to change the direction of the force). Students can also allow the ship to shoot a projectile using the \texttt{drawPoint()} function and the code includes an \texttt{if} statement that detects if spacebar is pressed for this purpose. This latter task is more difficult than the others because the projectile must be launched in the same direction as the ship whereas the planetoids can be given a random velocity using the \texttt{random()} function. One should also include the velocity of the ship when determining the velocity of the projectile as a fun illustration of Galilean invariance. Most students will just implement the reverse thrusters challenge.

\begin{figure}
    \centering
    \includegraphics[width=3.3in]{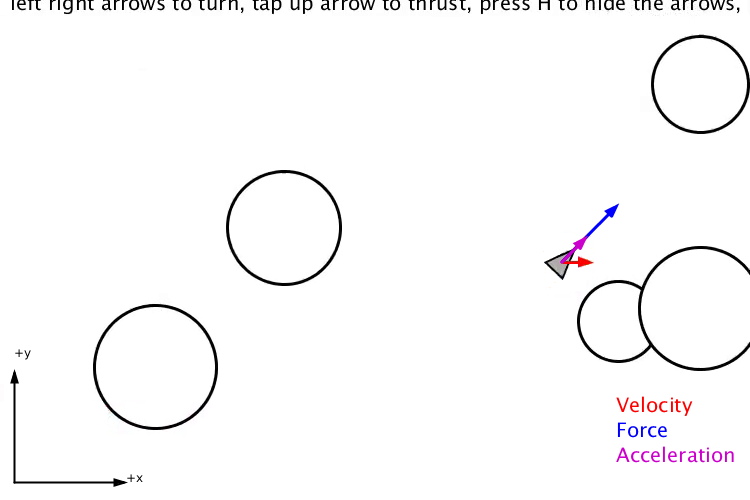}
    \caption{A screenshot from an activity where the student explores how changing the force of the rocket's thrust and the mass of the ship affects one's ability to avoid randomly drifting ``planetoids". This follows code modification tasks that enable the ship to move in two dimensions (instead of one dimension as in Fig.~\ref{fig:planetoids}).}
    \label{fig:withplanetoids}
\end{figure}

A more recent modification to this activity that was added after the study is to give the student a code that includes a number of drifting planetoids that will cause the game to end if the ship runs into one of them (Fig.~\ref{fig:withplanetoids}). The student is asked to explore the effect of changing the force of the ship's thrust and the mass of the ship on surviving in the game. This task helps foster a discussion of how it is only the ratio of the force to the mass that matters to the acceleration of a rocket in free space.

\begin{figure*}
\begin{center}
    \includegraphics[width=3.4in]{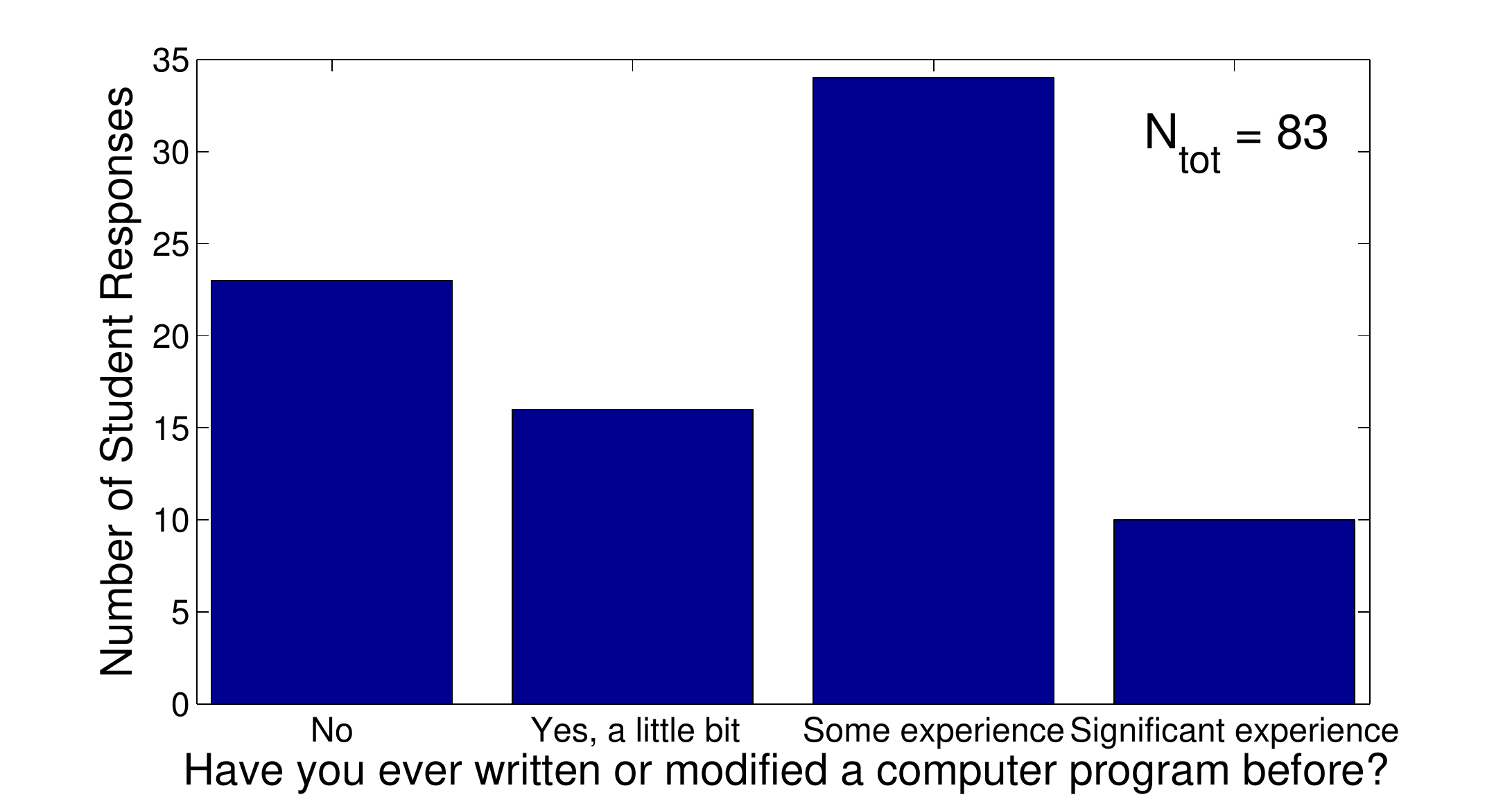}\includegraphics[width=3.4in]{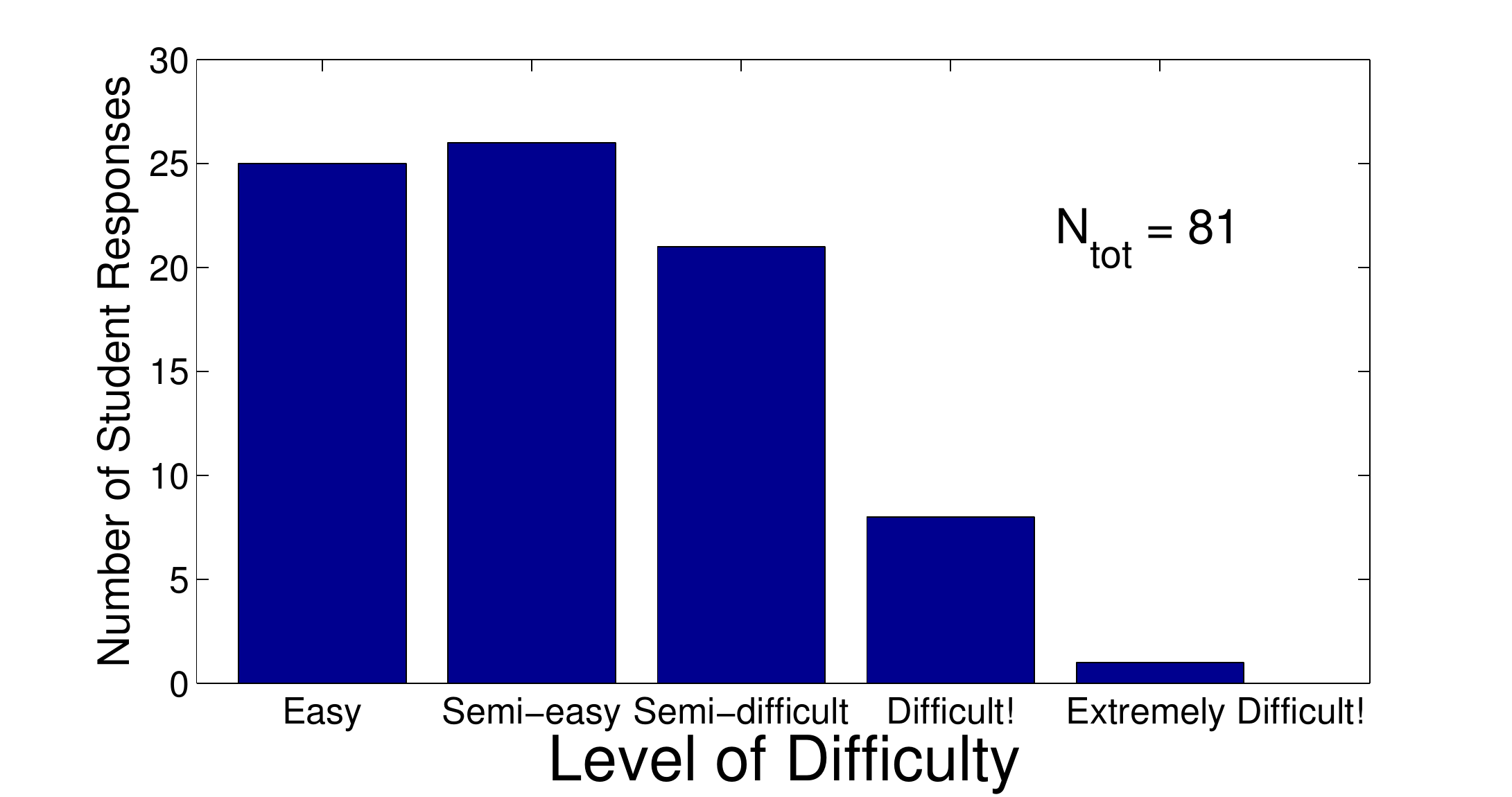}
    \includegraphics[width=3.4in]{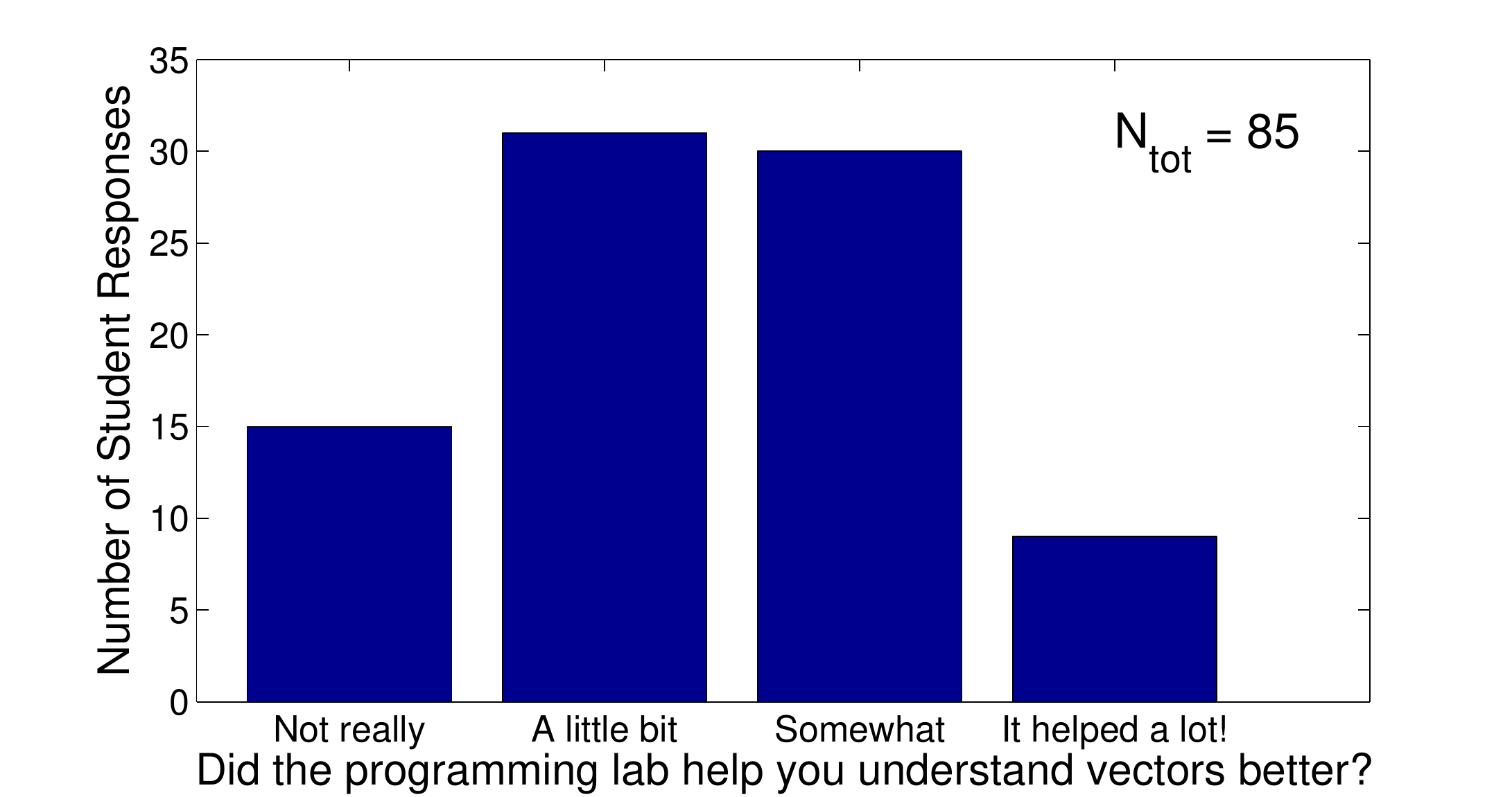}\includegraphics[width=3.4in]{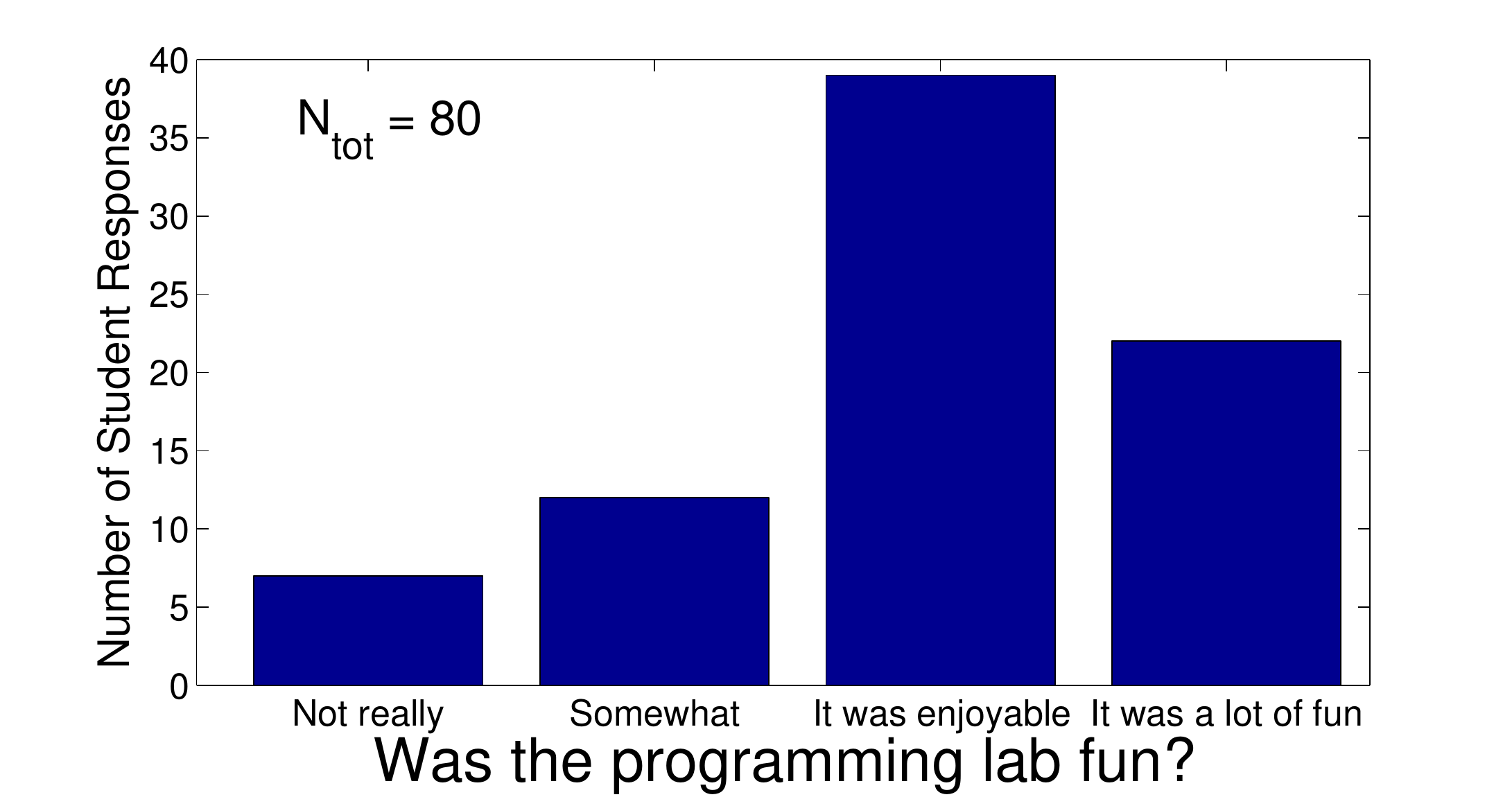}
    \end{center}
    \vspace{-0.6cm}
    \caption{Survey results from Ohio State Marion students who completed the first programming exercise (rocket in free space). Results are cumulative from four semesters of students (Spring 2015 - Fall 2016).} 
    \label{fig:survey}
\end{figure*}

\section{Student data}
\label{sec:survey}

After the student completes the Planetoids exercise, there is a detailed online survey that probes their experience in completing the activity. While the questions in this survey are qualitative and involve student self-reporting, the results can offer insight on whether the level of difficulty of the first exercise is appropriate and whether students find the exercises to be enjoyable to complete. 
Figure~\ref{fig:survey} summarizes the results of the survey from four semesters of students (Spring 2015 -- Fall 2016). The upper left plot in Fig.~\ref{fig:survey} shows that there are a significant number of absolute beginner programmers and weak programmers in the class. There were also a significant number of students who reported ``some experience" which may have meant that they were currently enrolled in a required C++ course, but had not had significant experience with coding prior to this.

The upper right plot in Fig.~\ref{fig:survey} shows that the difficulty level seems to be appropriate for the population of students, with a significant number of students selecting ``Easy!". The lower right plot in Fig.~\ref{fig:survey} indicates that many of the students found the programming activities to be enjoyable or fun. Students also have many positive things to say about the programming exercises in written evaluations at the end of the course after all of the exercises have been completed.

The bottom left plot in Fig.~\ref{fig:survey} summarizes student responses to the question ``Did the programming lab help you understand vectors better?" Although students can only provide a subjective estimation for how much they have learned, studies have shown that information of this kind can be valuable and even predictive other measures of student success \cite{Sawtelle_etal2012}. 

\section{Success Relative to Learning Goals}

In an earlier section (\ref{sec:learning}) we outlined two learning goals for the exercise. The student survey data in \ref{sec:survey} can provide some qualitative or indirect insight on whether these goals were met. In particular, the ``Level of Difficulty" question, which is asked after the completion of the code, relates to the learning goal of ``Understand how to convert a simple 1D code into a 2D code" since this is the main activity of the exercise. Unfortunately we do not have precise data to pinpoint the perceived difficulty for the subset of students who reported the least prior programming experience. But with only 1 student reporting ``Extremely Difficult!" and 8 students reporting ``Difficult!" compared to the 39 students who reported either ``No" or ``a little bit" of prior programming experience, the data supports the idea that students were able to complete the 1D to 2D conversion of the code without severe difficulty. Whether they fully understand the changes that were made is another important question that we can probe in future work.

The other learning goal was ``Understand how force, velocity and acceleration vectors relate to the motion of a ship traveling in free space". Although we do not have a direct probe of this learning goal, the question ``Did the programming lab help you understand vectors better?" relates to this learning objective in an indirect way.  Many of the students found the exercise to be at least ``somewhat" helpful in understanding vectors. As mentioned in the last section, student self-reporting can be useful and even predictive of student learning \cite{Sawtelle_etal2012}. In retrospect, one wonders if even more students would have reported understanding vectors better if there had been a part of the exercise where the student gives the ship an initial velocity and interacts with the program from that starting point, or if we had included the activity described earlier where students change the force (thrust) and mass of the ship (Fig.~\ref{fig:withplanetoids}) to see the effect on the motion in avoiding asteroids (an activity which was only added later).
It is also key to note that learning gains can only be achieved if students do actually engage with the activity. The question most closely related to this was ``Was the programming lab fun?" An overwhelming majority of the students found the exercise to be ``enjoyable" or ``fun" which suggests that they did significantly play around with the simulation (which demonstrates the relationship between force, velocity and acceleration vectors in an interactive way, making it very relevant to the goal of better understanding these vectors). It is therefore reasonable that there may be sufficiently high student "buy-in" to warrant further study, and further optimization of the user interface to maximize learning outcomes as described above.
Nevertheless, the questions discussed here are still oblique, self-reported measures of student learning on these learning goals and we do not wish to overstate the results we obtained. 

In future work we can directly probe the second learning goal using, for example, the rocket questions from either the Animated Force Concept Inventory by \citet{Dancy2006} or the conventional Force Concept Inventory \cite{FCI}, and other questions that ask students to identify the correct force, velocity and acceleration vectors in different situations. Importantly, we can compare results for these questions from students who complete a coding activity, and a ``control group" of students who only play around with the interactive for that coding activity for some period of time but without actually seeing or modifying the code. This will probe whether coding activities of the kind we discuss here, which involves multiple steps where students modify the code and check the behavior of the program, cause students to look more critically at the interactives they produce than they would if they did not have to perform coding tasks.

\section{Summary and Conclusion}

In this paper we illustrate a ``hybrid" approach to incorporating computer programming activities into introductory physics courses by describing a coding activity that resembles the classic asteroids game. The approach is so named because activities like the one described here produce interactives that bear some resemblance to web interactives that groups like PhET and Physlet have produced, but unlike PhET and Physlet, the student works with and modifies the code that evolves the system. In a ``traditional" computational physics course the student would have a great deal of control over producing visualizations. To reduce the cognitive load for weak or absolute beginner programmers in our study, the parts of the code that are unrelated to physics are hidden away in a \texttt{display()} function so that the student sees and works with only about 50 lines of code. In this sense our approach is a kind of ``hybrid" between canned interactives and mature computational physics exercises that are typically used in physics-major courses.

The first exercise in our suite of activities is an interactive simulation that resembles the classic game ``asteroids". The learning goals of this activity are to (1) understand how to convert a simple 1D code into a 2D code and (2) to understand how force, velocity and acceleration vectors relate to the motion of a ship traveling in free space. The activity includes scaffolding and hints to make the task of modifying the 1D code into 2D more manageable.

In an introductory class at OSU Marion where a substantial fraction of the students are weak or absolute beginner programmers, student survey data ($N \approx 80-85$) confirms that most students, including those with weak or absolute beginner programming experience, are able to complete the activity without severe difficulties. We interpret this as evidence that the first learning goal is being met. 

We are still only just beginning to investigate the effectiveness of the second learning goal. We discuss survey results that provide some insight into student experiences with the exercises, which in an indirect way addresses the second learning goal. However, this is no substitute for directly probing student learning with carefully chosen questions. In future work we will use the Animated Force Concept Inventory \cite{Dancy2006}, and other assessments to probe whether students understand the relationship between velocity and acceleration vectors. Of particular importance is whether the task of making modifications to the code and checking for the effect of these modifications on the interactive program will cause students to think more critically about the physics concepts than they would by playing around with a ``canned" interactive. This may be the real value of integrating coding at this level.   



We welcome inquiries from educators who may wish to use this suite of coding activities in their courses. Individual exercises and solution sets (including the planetoids game described here) are available at \url{http://compadre.org/PICUP}

\acknowledgements

The authors thank Chris Britt and Michael Hardesty for their collaboration on a p5.js learning management system. Chris Orban thanks Kathy Harper, Gregory Ngirmang, and Kelly Roos for discussions. This project was made possible through a Connect and Collaborate Grant, a program supporting innovative and scholarly engagement programs that leverage academic excellence of The Ohio State University in mutually beneficial ways with external partners. Support also comes from the American Institute of Physics Meggers Award.

\bibliographystyle{apsrev}
\bibliography{main}

\begin{thebibliography}{30}
\expandafter\ifx\csname natexlab\endcsname\relax\def\natexlab#1{#1}\fi
\expandafter\ifx\csname bibnamefont\endcsname\relax
  \def\bibnamefont#1{#1}\fi
\expandafter\ifx\csname bibfnamefont\endcsname\relax
  \def\bibfnamefont#1{#1}\fi
\expandafter\ifx\csname citenamefont\endcsname\relax
  \def\citenamefont#1{#1}\fi
\expandafter\ifx\csname url\endcsname\relax
  \def\url#1{\texttt{#1}}\fi
\expandafter\ifx\csname urlprefix\endcsname\relax\def\urlprefix{URL }\fi
\providecommand{\bibinfo}[2]{#2}
\providecommand{\eprint}[2][]{\url{#2}}

\bibitem[{\citenamefont{{Dancy} and {Beichner}}(2006)}]{Dancy2006}
\bibinfo{author}{\bibfnamefont{M.~H.} \bibnamefont{{Dancy}}} \bibnamefont{and}
  \bibinfo{author}{\bibfnamefont{R.}~\bibnamefont{{Beichner}}},
  \bibinfo{journal}{Physical Review Special Topics Physics Education}
  \textbf{\bibinfo{volume}{2}}, \bibinfo{eid}{010104} (\bibinfo{year}{2006}).

\bibitem[{\citenamefont{Fuller}(2006)}]{Fuller2006}
\bibinfo{author}{\bibfnamefont{R.~G.} \bibnamefont{Fuller}},
  \bibinfo{journal}{Computing in Science Engineering}
  \textbf{\bibinfo{volume}{8}}, \bibinfo{pages}{16} (\bibinfo{year}{2006}),
  ISSN \bibinfo{issn}{1521-9615}.

\bibitem[{\citenamefont{of~Labor~Statistics}(2014)}]{labor2014}
\bibinfo{author}{\bibfnamefont{U.~B.} \bibnamefont{of~Labor~Statistics}},
  \bibinfo{journal}{Online Publication}  (\bibinfo{year}{2014}),
  \urlprefix\url{https://www.bls.gov/careeroutlook/2014/spring/art01.pdf}.

\bibitem[{\citenamefont{{Code.org}}()}]{code}
\bibinfo{author}{\bibnamefont{{Code.org}}}, \emph{\bibinfo{title}{Promote
  computer science}}, \bibinfo{howpublished}{\url{https://code.org/promote}},
  \bibinfo{note}{accessed: 2017-07-06}.

\bibitem[{\citenamefont{{Smith}}(2015)}]{coresubject}
\bibinfo{author}{\bibfnamefont{D.~F.} \bibnamefont{{Smith}}},
  \emph{\bibinfo{title}{What the essa means for the future of computer science
  in stem}},
  \bibinfo{howpublished}{\url{http://www.edtechmagazine.com/k12/article/2015/12/what-essa-means-future-computer-science-and-stem}}
  (\bibinfo{year}{2015}), \bibinfo{note}{accessed: 2016-12-30}.

\bibitem[{\citenamefont{{Chabay} and {Sherwood}}(2008)}]{Chabay_Sherwood2008}
\bibinfo{author}{\bibfnamefont{R.}~\bibnamefont{{Chabay}}} \bibnamefont{and}
  \bibinfo{author}{\bibfnamefont{B.}~\bibnamefont{{Sherwood}}},
  \bibinfo{journal}{American Journal of Physics} \textbf{\bibinfo{volume}{76}},
  \bibinfo{pages}{307} (\bibinfo{year}{2008}).

\bibitem[{\citenamefont{{Caballero} et~al.}(2012)\citenamefont{{Caballero},
  {Kohlmyer}, and {Schatz}}}]{Caballero_etal2012}
\bibinfo{author}{\bibfnamefont{M.~D.} \bibnamefont{{Caballero}}},
  \bibinfo{author}{\bibfnamefont{M.~A.} \bibnamefont{{Kohlmyer}}},
  \bibnamefont{and} \bibinfo{author}{\bibfnamefont{M.~F.}
  \bibnamefont{{Schatz}}}, \bibinfo{journal}{Physical Review Special Topics
  Physics Education} \textbf{\bibinfo{volume}{8}}, \bibinfo{eid}{020106}
  (\bibinfo{year}{2012}), \eprint{1107.5216}.

\bibitem[{\citenamefont{{Esquembre} and {Titus}}(2016)}]{Titus_Esquembre2016}
\bibinfo{author}{\bibfnamefont{F.}~\bibnamefont{{Esquembre}}} \bibnamefont{and}
  \bibinfo{author}{\bibfnamefont{A.}~\bibnamefont{{Titus}}},
  \emph{\bibinfo{title}{Exploring physics with video games}},
  \bibinfo{howpublished}{\url{http://www.opensourcephysics.org/items/detail.cfm?ID=13970}}
  (\bibinfo{year}{2016}), \bibinfo{note}{accessed: 2017-07-05}.

\bibitem[{\citenamefont{{Aiken} et~al.}(2013)\citenamefont{{Aiken},
  {Caballero}, {Douglas}, {Burk}, {Scanlon}, {Thoms}, and
  {Schatz}}}]{Aiken_etal2013}
\bibinfo{author}{\bibfnamefont{J.~M.} \bibnamefont{{Aiken}}},
  \bibinfo{author}{\bibfnamefont{M.~D.} \bibnamefont{{Caballero}}},
  \bibinfo{author}{\bibfnamefont{S.~S.} \bibnamefont{{Douglas}}},
  \bibinfo{author}{\bibfnamefont{J.~B.} \bibnamefont{{Burk}}},
  \bibinfo{author}{\bibfnamefont{E.~M.} \bibnamefont{{Scanlon}}},
  \bibinfo{author}{\bibfnamefont{B.~D.} \bibnamefont{{Thoms}}},
  \bibnamefont{and} \bibinfo{author}{\bibfnamefont{M.~F.}
  \bibnamefont{{Schatz}}}, in \emph{\bibinfo{booktitle}{American Institute of
  Physics Conference Series}}, edited by \bibinfo{editor}{\bibfnamefont{P.~V.}
  \bibnamefont{{Engelhardt}}}, \bibinfo{editor}{\bibfnamefont{A.~D.}
  \bibnamefont{{Churukian}}}, \bibnamefont{and}
  \bibinfo{editor}{\bibfnamefont{N.~S.} \bibnamefont{{Rebello}}}
  (\bibinfo{year}{2013}), vol. \bibinfo{volume}{1513} of
  \emph{\bibinfo{series}{American Institute of Physics Conference Series}}, pp.
  \bibinfo{pages}{46--49}, \eprint{1207.1764}.

\bibitem[{\citenamefont{{Aho} et~al.}(2014)\citenamefont{{Aho}, {Chandra}, and
  {Roberts}}}]{Aho2014}
\bibinfo{author}{\bibfnamefont{K.}~\bibnamefont{{Aho}}},
  \bibinfo{author}{\bibfnamefont{K.}~\bibnamefont{{Chandra}}},
  \bibnamefont{and}
  \bibinfo{author}{\bibfnamefont{E.}~\bibnamefont{{Roberts}}},
  \bibinfo{journal}{Proceedings of the American Society for Engineering
  Education}  (\bibinfo{year}{2014}).

\bibitem[{\citenamefont{Wieman et~al.}(2010)\citenamefont{Wieman, Adams,
  Loeblein, and Perkins}}]{PhET}
\bibinfo{author}{\bibfnamefont{C.~E.} \bibnamefont{Wieman}},
  \bibinfo{author}{\bibfnamefont{W.~K.} \bibnamefont{Adams}},
  \bibinfo{author}{\bibfnamefont{P.}~\bibnamefont{Loeblein}}, \bibnamefont{and}
  \bibinfo{author}{\bibfnamefont{K.~K.} \bibnamefont{Perkins}},
  \bibinfo{journal}{The Physics Teacher} \textbf{\bibinfo{volume}{48}},
  \bibinfo{pages}{225} (\bibinfo{year}{2010}),
  \eprint{http://dx.doi.org/10.1119/1.3361987},
  \urlprefix\url{http://dx.doi.org/10.1119/1.3361987}.

\bibitem[{\citenamefont{{Christian} and {Belloni}}(2003)}]{physlet}
\bibinfo{author}{\bibfnamefont{W.}~\bibnamefont{{Christian}}} \bibnamefont{and}
  \bibinfo{author}{\bibfnamefont{M.}~\bibnamefont{{Belloni}}},
  \emph{\bibinfo{title}{Physlet Physics: Interactive Illustrations,
  Explorations, and Problems for Introductory Physics}}
  (\bibinfo{publisher}{Addison-Wesley}, \bibinfo{year}{2003}).

\bibitem[{\citenamefont{{Perkins} et~al.}(2006)\citenamefont{{Perkins},
  {Adams}, {Dubson}, {Finkelstein}, {Reid}, {Wieman}, and
  {LeMaster}}}]{Perkins_etal2006}
\bibinfo{author}{\bibfnamefont{K.}~\bibnamefont{{Perkins}}},
  \bibinfo{author}{\bibfnamefont{W.}~\bibnamefont{{Adams}}},
  \bibinfo{author}{\bibfnamefont{M.}~\bibnamefont{{Dubson}}},
  \bibinfo{author}{\bibfnamefont{N.}~\bibnamefont{{Finkelstein}}},
  \bibinfo{author}{\bibfnamefont{S.}~\bibnamefont{{Reid}}},
  \bibinfo{author}{\bibfnamefont{C.}~\bibnamefont{{Wieman}}}, \bibnamefont{and}
  \bibinfo{author}{\bibfnamefont{R.}~\bibnamefont{{LeMaster}}},
  \bibinfo{journal}{The Physics Teacher} \textbf{\bibinfo{volume}{44}},
  \bibinfo{pages}{18} (\bibinfo{year}{2006}).

\bibitem[{\citenamefont{Podolefsky et~al.}(2010)\citenamefont{Podolefsky,
  Perkins, and Adams}}]{Podolefsky_etal2010}
\bibinfo{author}{\bibfnamefont{N.~S.} \bibnamefont{Podolefsky}},
  \bibinfo{author}{\bibfnamefont{K.~K.} \bibnamefont{Perkins}},
  \bibnamefont{and} \bibinfo{author}{\bibfnamefont{W.~K.} \bibnamefont{Adams}},
  \bibinfo{journal}{Phys. Rev. ST Phys. Educ. Res.}
  \textbf{\bibinfo{volume}{6}}, \bibinfo{pages}{020117} (\bibinfo{year}{2010}),
  \urlprefix\url{http://link.aps.org/doi/10.1103/PhysRevSTPER.6.020117}.

\bibitem[{\citenamefont{Jong}(2010)}]{Jong_2010}
\bibinfo{author}{\bibfnamefont{T.~d.} \bibnamefont{Jong}},
  \bibinfo{journal}{Instructional Science} \textbf{\bibinfo{volume}{38}},
  \bibinfo{pages}{105} (\bibinfo{year}{2010}),
  \urlprefix\url{http://dx.doi.org/10.1007/s11251-009-9110-0}.

\bibitem[{\citenamefont{{Mayer} and {Moreno}}(2003)}]{cogload}
\bibinfo{author}{\bibfnamefont{R.~E.} \bibnamefont{{Mayer}}} \bibnamefont{and}
  \bibinfo{author}{\bibfnamefont{R.}~\bibnamefont{{Moreno}}},
  \bibinfo{journal}{Educational Psychology} \textbf{\bibinfo{volume}{38}},
  \bibinfo{pages}{43} (\bibinfo{year}{2003}).

\bibitem[{\citenamefont{Kordaki}(2010)}]{kordakai2010}
\bibinfo{author}{\bibfnamefont{M.}~\bibnamefont{Kordaki}},
  \bibinfo{journal}{Computers \& Education} \textbf{\bibinfo{volume}{54}},
  \bibinfo{pages}{69 } (\bibinfo{year}{2010}), ISSN \bibinfo{issn}{0360-1315},
  \urlprefix\url{http://www.sciencedirect.com/science/article/pii/S0360131509001845}.

\bibitem[{\citenamefont{{Tisue} and {Wilensky}}(2004)}]{netlogo}
\bibinfo{author}{\bibfnamefont{S.}~\bibnamefont{{Tisue}}} \bibnamefont{and}
  \bibinfo{author}{\bibfnamefont{U.}~\bibnamefont{{Wilensky}}}, in
  \emph{\bibinfo{booktitle}{In Proceedings of Agent 2004}}
  (\bibinfo{year}{2004}).

\bibitem[{\citenamefont{Taub et~al.}(2015)\citenamefont{Taub, Armoni, Bagno,
  and Ben-Ari}}]{Taub_etal2015}
\bibinfo{author}{\bibfnamefont{R.}~\bibnamefont{Taub}},
  \bibinfo{author}{\bibfnamefont{M.}~\bibnamefont{Armoni}},
  \bibinfo{author}{\bibfnamefont{E.}~\bibnamefont{Bagno}}, \bibnamefont{and}
  \bibinfo{author}{\bibfnamefont{M.~M.} \bibnamefont{Ben-Ari}},
  \bibinfo{journal}{Computers \& Education} \textbf{\bibinfo{volume}{87}},
  \bibinfo{pages}{10 } (\bibinfo{year}{2015}), ISSN \bibinfo{issn}{0360-1315},
  \urlprefix\url{http://www.sciencedirect.com/science/article/pii/S0360131515000913}.

\bibitem[{\citenamefont{Weintrop et~al.}(2016)\citenamefont{Weintrop, Beheshti,
  Horn, Orton, Jona, Trouille, and Wilensky}}]{Weintrop_etal2016}
\bibinfo{author}{\bibfnamefont{D.}~\bibnamefont{Weintrop}},
  \bibinfo{author}{\bibfnamefont{E.}~\bibnamefont{Beheshti}},
  \bibinfo{author}{\bibfnamefont{M.}~\bibnamefont{Horn}},
  \bibinfo{author}{\bibfnamefont{K.}~\bibnamefont{Orton}},
  \bibinfo{author}{\bibfnamefont{K.}~\bibnamefont{Jona}},
  \bibinfo{author}{\bibfnamefont{L.}~\bibnamefont{Trouille}}, \bibnamefont{and}
  \bibinfo{author}{\bibfnamefont{U.}~\bibnamefont{Wilensky}},
  \bibinfo{journal}{Journal of Science Education and Technology}
  \textbf{\bibinfo{volume}{25}}, \bibinfo{pages}{127} (\bibinfo{year}{2016}),
  ISSN \bibinfo{issn}{1573-1839},
  \urlprefix\url{http://dx.doi.org/10.1007/s10956-015-9581-5}.

\bibitem[{\citenamefont{{Chabay} and {Sherwood}}(2015)}]{Chabay_Sherwood2015}
\bibinfo{author}{\bibfnamefont{R.}~\bibnamefont{{Chabay}}} \bibnamefont{and}
  \bibinfo{author}{\bibfnamefont{B.}~\bibnamefont{{Sherwood}}},
  \emph{\bibinfo{title}{{Matter \& Interactions, 4th edition}}}
  (\bibinfo{publisher}{Wiley \& Sons}, \bibinfo{year}{2015}).

\bibitem[{\citenamefont{{Serbanescu} et~al.}(2011)\citenamefont{{Serbanescu},
  {Kushner}, and {Stanley}}}]{Serbanescu_etal2011}
\bibinfo{author}{\bibfnamefont{R.~M.} \bibnamefont{{Serbanescu}}},
  \bibinfo{author}{\bibfnamefont{P.~J.} \bibnamefont{{Kushner}}},
  \bibnamefont{and}
  \bibinfo{author}{\bibfnamefont{S.}~\bibnamefont{{Stanley}}},
  \bibinfo{journal}{American Journal of Physics} \textbf{\bibinfo{volume}{79}},
  \bibinfo{pages}{919} (\bibinfo{year}{2011}).

\bibitem[{\citenamefont{Marion}(2014)}]{osumarion2014}
\bibinfo{author}{\bibfnamefont{O.}~\bibnamefont{Marion}},
  \emph{\bibinfo{title}{Quick facts, the ohio state university at marion}},
  \bibinfo{howpublished}{\url{http://osumarion.osu.edu/about/quick-facts.html}}
  (\bibinfo{year}{2014}), \bibinfo{note}{accessed: 2014-1-1}.

\bibitem[{\citenamefont{Marion}(2015)}]{osumarion2015}
\bibinfo{author}{\bibfnamefont{O.}~\bibnamefont{Marion}},
  \emph{\bibinfo{title}{Quick facts, the ohio state university at marion}},
  \bibinfo{howpublished}{\url{http://osumarion.osu.edu/about/quick-facts.html}}
  (\bibinfo{year}{2015}), \bibinfo{note}{accessed: 2015-1-1}.

\bibitem[{\citenamefont{Marion}(2016)}]{osumarion2016}
\bibinfo{author}{\bibfnamefont{O.}~\bibnamefont{Marion}},
  \emph{\bibinfo{title}{Quick facts, the ohio state university at marion}},
  \bibinfo{howpublished}{\url{http://osumarion.osu.edu/about/quick-facts.html}}
  (\bibinfo{year}{2016}), \bibinfo{note}{accessed: 2016-1-1}.

\bibitem[{\citenamefont{Columbus}(2017)}]{osucolumbus}
\bibinfo{author}{\bibfnamefont{O.}~\bibnamefont{Columbus}},
  \emph{\bibinfo{title}{2017 enrollment report}},
  \bibinfo{howpublished}{\url{http://enrollmentservices.osu.edu/report.pdf}}
  (\bibinfo{year}{2017}), \bibinfo{note}{accessed: 2017-10-29}.

\bibitem[{\citenamefont{Cromer}(1981)}]{Cromer1981}
\bibinfo{author}{\bibfnamefont{A.}~\bibnamefont{Cromer}},
  \bibinfo{journal}{American Journal of Physics} \textbf{\bibinfo{volume}{49}},
  \bibinfo{pages}{455} (\bibinfo{year}{1981}),
  \eprint{http://dx.doi.org/10.1119/1.12478},
  \urlprefix\url{http://dx.doi.org/10.1119/1.12478}.

\bibitem[{\citenamefont{White}(1984)}]{white1984}
\bibinfo{author}{\bibfnamefont{B.~Y.} \bibnamefont{White}},
  \bibinfo{journal}{Cognition and Instruction} \textbf{\bibinfo{volume}{1}},
  \bibinfo{pages}{69} (\bibinfo{year}{1984}),
  \urlprefix\url{https://doi.org/10.1207/s1532690xci0101_4}.

\bibitem[{\citenamefont{Sawtelle et~al.}(2012)\citenamefont{Sawtelle, Brewe,
  and Kramer}}]{Sawtelle_etal2012}
\bibinfo{author}{\bibfnamefont{V.}~\bibnamefont{Sawtelle}},
  \bibinfo{author}{\bibfnamefont{E.}~\bibnamefont{Brewe}}, \bibnamefont{and}
  \bibinfo{author}{\bibfnamefont{L.~H.} \bibnamefont{Kramer}},
  \bibinfo{journal}{Journal of Research in Science Teaching}
  \textbf{\bibinfo{volume}{49}}, \bibinfo{pages}{1096} (\bibinfo{year}{2012}),
  ISSN \bibinfo{issn}{1098-2736},
  \urlprefix\url{http://dx.doi.org/10.1002/tea.21050}.

\bibitem[{\citenamefont{{Hestenes} et~al.}(1992)\citenamefont{{Hestenes},
  {Wells}, and {Swackhamer}}}]{FCI}
\bibinfo{author}{\bibfnamefont{D.}~\bibnamefont{{Hestenes}}},
  \bibinfo{author}{\bibfnamefont{M.}~\bibnamefont{{Wells}}}, \bibnamefont{and}
  \bibinfo{author}{\bibfnamefont{G.}~\bibnamefont{{Swackhamer}}},
  \bibinfo{journal}{The Physics Teacher} \textbf{\bibinfo{volume}{30}},
  \bibinfo{pages}{141} (\bibinfo{year}{1992}).

\end{thebibliography}

\end{document}